\documentclass[twocolumn,aps,prd,superscriptaddress,amsmath,amssymb,nofootinbib,nobibnotes]{revtex4-2}
	\usepackage{graphicx} 
	\usepackage{amsfonts, times, amsthm}
	\usepackage{bm} 
    \usepackage[normalem]{ulem}
	\usepackage{epstopdf}
	\usepackage[linktocpage,colorlinks=true,urlcolor=blue]{hyperref}
	\usepackage[dvipsnames]{xcolor}     
	\usepackage{natbib}
	\usepackage{soul}
    \usepackage{subfigure}
    \usepackage{hhline}
	\usepackage[utf8x]{inputenc}
	\definecolor{darkgreen}{rgb}{0,0.5,0}

	\def\p{\partial}
	
	\newcommand{\sss}{\scriptscriptstyle}

\def\KN{\text{KN}}
\def\KS{\text{KS}}

	\DeclareUnicodeCharacter{2009}{\,} 
	\begin{document}
		\title{Degeneracies in the quasinormal mode spectrum of the Kerr-Sen black hole}

			\author{Sérgio V.~M.~C.~B.~Xavier}
		\email{sergio.xavier@ufabc.edu.br}
		\affiliation{Centro de Matemática, Computação e Cognição, \\Universidade Federal do ABC (UFABC), 09210-580, Santo André, São Paulo, Brazil}
	
     \author{João Paulo Cavalcante}
	\email{joao.cavalcante@ufabc.edu.br}
	\affiliation{Centro de Matemática, Computação e Cognição, \\Universidade Federal do ABC (UFABC), 09210-580, Santo André, São Paulo, Brazil}

 \author{Enzo Boschilia Vita}
	\email{enzo.vita@ufabc.edu.br}
	\affiliation{Centro de Matemática, Computação e Cognição, \\Universidade Federal do ABC (UFABC), 09210-580, Santo André, São Paulo, Brazil}
 
      \author{Maurício Richartz}
	\email{mauricio.richartz@ufabc.edu.br}
	\affiliation{Centro de Matemática, Computação e Cognição, \\Universidade Federal do ABC (UFABC), 09210-580, Santo André, São Paulo, Brazil}

\begin{abstract}
		The non-Hermitian nature of the eigenvalue problem underlying black hole perturbations can induce the presence of exceptional points (EPs) in the quasinormal mode (QNM) spectrum. EPs are points in parameter space where QNMs become degenerate. Near an EP, QNMs are resonantly excited, leaving imprints on the ringdown signal. In this work, we investigate the exceptional structure of scalar perturbations of the Kerr-Sen black hole, a rotating, charged solution of the low-energy limit of heterotic string theory, and compare it with the EP structure of the Kerr-Newman black hole in General Relativity. Starting from the EP associated with a massive scalar field in the Kerr limit, we trace exceptional lines in the parameter space spanned by the black hole spin, charge, and field mass. We locate these degeneracies directly by imposing double-root conditions on the QNM secular equation, thus avoiding the need for parameter space scans. We also investigate the hysteresis phenomenon, showing that encircling an EP in parameter space leads to an exchange of the associated QNMs. 
\end{abstract}

\date{\today}

\maketitle

\section{Introduction}\label{sec:int}

Since the first detection of gravitational waves in 2015~\cite{LIGOScientific:2016aoc}, the LIGO-Virgo-KAGRA collaboration has observed hundreds of events~\cite{LIGOScientific:2026sit}, allowing gravity to be tested in the strong-field regime~\cite{LIGOScientific:2026qni,Yunes:2024lzm}. Many more detections are expected from LISA~\cite{Barausse:2020rsu} and from next-generation ground-based detectors such as the Einstein Telescope and Cosmic Explorer~\cite{ET:2025xjr,Reitze:2019iox}. Interpreting these observations requires an accurate description of the post-merger stage of compact binary coalescences. The standard approach to model the ringdown stage relies on black hole (BH) perturbation theory~\cite{Berti:2025hly,Franchini:2023eda}, in which the response of the remnant BH is described by a superposition of damped oscillations, called quasinormal modes (QNMs)~\cite{Nollert:1999ji,Konoplya:2011qq,Berti:2009kk,Bolokhov:2025uxz}.

QNM frequencies are complex numbers, with the real part giving the oscillation frequency and the imaginary part the decay rate. Since they depend only on the background spacetime and on the field equations governing the perturbation, QNMs are regarded as fingerprints of BHs. BH spectroscopy therefore provides a direct test of General Relativity and of the Kerr hypothesis, i.e., the idea that astrophysical dark compact objects are described by the Kerr metric~\cite{Herdeiro:2022yle}. In particular, it can determine whether these objects are fully characterized by their mass and angular momentum or harbor additional fundamental fields~\cite{Berti:2025hly,Cardoso:2016rao,Cardoso:2019rvt,Barack:2018yly,LIGOScientific:2021sio,LIGOScientific:2026wpt,LIGOScientific:2026oim}.

The damped character of QNMs reveals that BH spacetimes are open, dissipative systems. Waves are either absorbed by the horizon or radiated to infinity, and the energy of the perturbation is not conserved. Therefore, the QNM problem is intrinsically non-Hermitian. Non-Hermitian physics is an active area of research, with a well-developed theory in optics, photonics, condensed matter, and quantum mechanics~\cite{Ashida:2020dkc,Bergholtz:2019deh,Ding:2022juv,ElGanainy:2018,Ozdemir:2019,Miri:2019}. A distinctive feature of non-Hermitian systems is the presence of exceptional points (EPs), spectral degeneracies at which two or more eigenvalues and their eigenfunctions coalesce~\cite{Ding:2022juv,Heiss_2012}.
Near an EP, the eigenvalue spectrum has a nontrivial structure in parameter space. For a second-order EP, varying the system parameters along a closed loop encircling the EP interchanges the two associated eigenvalues, such that two complete circuits are required to recover the original labeling~\cite{Heiss_2012,Dembowski:2001,Berti:2025hly}. More generally, the number of circuits required depends on the permutation of the modes induced by the loop. Consequently, the labeling of the modes depends on the path taken in parameter space, a phenomenon known as hysteresis~\cite{Cavalcante:2024swt,Cavalcante:2025abr}.

Recently, the existence of EPs in the QNM spectrum of BHs was reported. In Ref.~\cite{Motohashi:2024fwt}, Motohashi showed that the avoided crossing observed in the Kerr QNM spectrum occurs in the vicinity of an EP and is accompanied by a resonant excitation of the modes involved. Similar eigenvalue repulsions had previously been reported in the Kerr-Newman (KN) spectrum~\cite{Dias:2021yju,Dias:2022}. Independently, the authors of Refs.~\cite{Cavalcante:2024swt,Cavalcante:2025abr} located an EP in the spectrum of massive scalar perturbations of near-extremal Kerr BHs and showed that encircling it leads to hysteresis. Since then, EP structures have been found in rotating and charged spacetimes~\cite{Cavalcante:2026vgr,Nakamoto:2026lyo,Rossi:2026als,Destounis:2026qne,Arcos:2026eis,Zhao:2026nce}, under localized modifications of the effective potential~\cite{OuldElHadj:2026vym,Wu:2026tfx}, in coupled perturbation systems~\cite{Takahashi:2025uwo}, and in hairy BHs~\cite{Cheng:2026gxu}. The resonant enhancement of QNM excitation near EPs, together with its potential observational signatures in the ringdown signal, has also been discussed in Refs.~\cite{Oshita:2025ibu,Yang:2025dbn,Takahashi:2025uwo,Kubota:2026hdv,PanossoMacedo:2025xnf,Rossi:2026als,Takahashi:2026daf,Long:2026vpw,Imafuku:2026rpn,Zhang:2026uoo}.

At an EP, the simple poles of the Green's function associated with the QNMs merge into a higher-order pole~\cite{PanossoMacedo:2025xnf,Morikawa:2026fzn}. EPs typically occur at isolated points in a two-dimensional parameter space, such as the one spanned by the Kerr spin and the scalar field mass~\cite{Cavalcante:2024swt}. Introducing a third parameter allows these isolated points to extend into exceptional lines, as recently investigated for KN BHs~\cite{Cavalcante:2026vgr}, see also Refs.~\cite{Nakamoto:2026lyo,Cao:2025afs}. However, how the shape and endpoints of these lines depend on the underlying spacetime geometry remains largely unexplored. 
To investigate this dependence, we extend the analysis of exceptional lines beyond Einstein--Maxwell theory. We focus on the Kerr-Sen BH (KSBH), a rotating and electrically charged solution of the low-energy limit of heterotic string theory~\cite{Sen:PRL1992}, and compare our results with those for the KN solution of General Relativity~\cite{Cavalcante:2026vgr}. 

Although both the KS and the KN spacetimes are characterized by their mass $M$, charge $Q$, and angular momentum per unit mass $a$, and share the same asymptotic behavior, they differ in the strong-field regime~\cite{Burinskii:1995,Delgado_Herdeiro_Radu:2016}. Starting from EPs associated with a massive scalar field in the Kerr limit, we trace exceptional lines by imposing double-root conditions on the QNM secular equation, thus locating mode degeneracies directly without scanning the parameter space. In both the KNBH and KSBH spacetimes, the exceptional lines extend towards lower scalar field masses as the charge increases, approaching extremality but terminating at zero scalar field mass before reaching it. We also investigate the relation between EPs and the bifurcation of the spectrum into damped modes (DMs) and zero-damping modes (ZDMs)~\cite{Yang:2013uba,Kokkotas:2015uma}, and explore the associated hysteresis phenomenon.

The remainder of this paper is organized as follows. In Sec.~\ref{sec:2}, we present the KSBH spacetime, compare its parameter space with that of the KNBH, and separate the Klein-Gordon equation for a massive scalar field. In Sec.~\ref{sec:3}, we formulate the QNM problem, describe the continued fraction method used to compute the spectrum and introduce the double-root conditions that locate EPs and allow exceptional lines to be traced. In Sec.~\ref{sec:4}, we present the exceptional lines for KS and KN BHs and discuss the EPs that emerge in the massless limit. We then investigate the relation between these EPs and the bifurcation of the spectrum into DMs and ZDMs, as well as the associated mode exchange and hysteresis. Our final remarks are presented in Sec.~\ref{sec:remarks}. We adopt natural units $G=c=\hbar=1$.

\section{Scalar field on a stringy BH spacetime}\label{sec:2}

The KSBH is a solution of the low-energy regime of heterotic string theory, in which the energy-matter sector of the action is formed by a dilaton, an axion, and a $U(1)$ gauge field~\cite{Sen:PRL1992}. In the Einstein frame and in Boyer-Lindquist coordinates $(t,r,\theta,\varphi)$, its line element reads
	\begin{align}
		ds^{2}=&-\left(1-\frac{2 M r}{\rho^{2}}\right)dt^{2}+\rho^{2}\left(\frac{dr^{2}}{\Delta}+d\theta^{2}\right)-\nonumber\\
        & \hspace{-0.6cm} \frac{4Mra\sin^{2}\theta}{\rho^{2}}dtd\varphi+\!\left(\!\xi+\frac{2 M r a^{2}\sin^{2}\theta}{\rho^{2}}\right) \! \sin^{2} \!\theta\,d\varphi^{2},
		\label{ds2KS}
	\end{align}
where 
	\begin{align}
		&\Delta = r\left(r+2d\right)-2 M r+a^{2},\\
		&\rho^{2} = r\left(r+2d\right)+a^{2}\cos^{2}\theta,\\
		&\xi = r\left(r+2d\right)+a^{2},
	\end{align}
with 
\begin{equation}
d = \frac{Q^2}{2M}. \label{d}
\end{equation}
The parameters $(M, Q, a)$ represent, respectively, the mass, the electric charge and the angular momentum per unit mass of the KSBH. Additionally, the inner ($r_{-}^{\sss \text{KS}}$) and outer ($r_{+}^{\sss \text{KS}}$) horizons of the KSBH are located at
	\begin{equation}
		r_{\pm}^{\sss \text{KS}} = M-d\pm\sqrt{\left(M-d\right)^2-a^2}. 
		\label{rh_ks}
	\end{equation}

\begin{figure}
    \centering
    \includegraphics[width=1\linewidth]{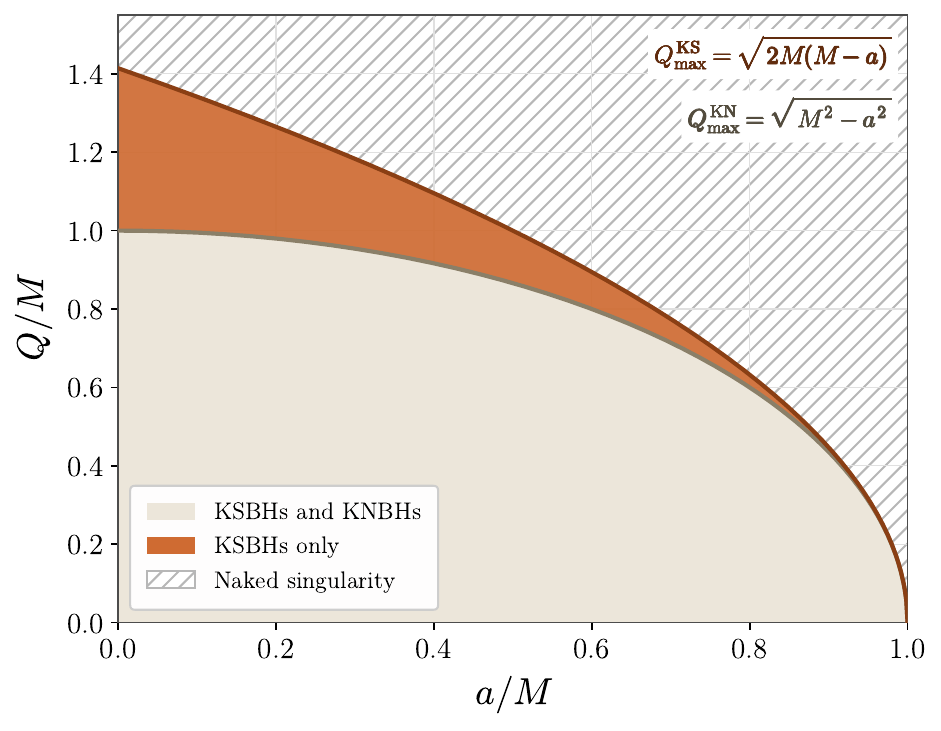}
    \caption{Parameter space of KSBHs and KNBHs in the $(a/M,\,Q/M)$ plane. The lighter (beige) region corresponds to configurations where both KSBHs and KNBHs exist, while the darker (orange) region is accessible only to KSBHs. The hatched region corresponds to naked singularities in both spacetimes. The solid curves indicate the extremal boundaries of the two BH families.}
    \label{ChargeKNvsKS}
\end{figure}

To assess how the exceptional structure depends on the underlying theory of gravity, we compare the KSBH with the KNBH, the corresponding rotating, charged solution of General Relativity. The line element of the KNBH, in Boyer-Lindquist coordinates, is ~\cite{Newman:1965my,Wald:1984}
		\begin{align}
	ds^{2}=&-\left(1-\frac{2 M r - Q^{2}}{\rho^{2}_{\KN}}\right)dt^{2}+\rho^{2}_{\KN}\left(\frac{dr^{2}}{\Delta_{\KN}}+d\theta^{2}\right)\nonumber\\
	&-\frac{4 M a r\sin^{2}\theta-2a Q^{2}\sin^{2}\theta}{\rho^{2}_{\KN}}dtd\varphi\nonumber\\
	&+\frac{\left[(r^{2}+a^{2})^{2}-\Delta_{\KN} a^{2}\sin^{2}\theta\right]\sin^{2}\theta}{\rho^{2}_{\KN}}d\varphi^{2},
	\label{ds2KN}
	\end{align}	
	where
	\begin{align}
	&\Delta_{\text{KN}} = r^{2}-2 M r+a^{2}+Q^{2},\\
	&\rho^{2}_{\text{KN}} = r^{2}+a^{2}\cos^{2}\theta.
	\end{align}
The inner ($r_{-}^{\sss \text{KN}}$) and outer ($r_{+}^{\sss \text{KN}}$) horizons of the KNBH are located at the following radial coordinates,
	\begin{equation}
	r_{\pm}^{\sss \KN} = M \pm\sqrt{M^{2}-a^{2}-Q^{2}}.
	\label{rh_kn}
	\end{equation}

Despite their similarities, the KSBH and the KNBH exhibit important differences~\cite{Burinskii:1995,Xavier:2020egv}. In particular, the KSBH can sustain a larger electric charge before reaching extremality. This difference arises from the additional term defined in Eq.~\eqref{d}, which shifts the horizon location in the KSBH (see Eq.~\eqref{rh_ks}) relative to the KN case (see Eq.~\eqref{rh_kn}). The maximum allowed charges (for a given spin $a<M$) are
    \begin{align}
        &Q_{\text{max}}^{\sss \KS}=\sqrt{2M(M-a)},\\
        &Q_{\text{max}}^{\sss \KN}=\sqrt{M^2-a^2}.
    \end{align}
In the nonrotating limit, the KSBH reduces to the Gibbons–Maeda–Garfinkle–Horowitz–Strominger BH~\cite{Gibbons:1987ps,Garfinkle:1990qj}, whose maximum charge exceeds the Reissner–Nordström extremal bound $Q=M$ by a factor of $\sqrt{2}$. Figure~\ref{ChargeKNvsKS} compares the allowed parameter spaces of the two BHs, with the orange region corresponding to configurations accessible only to the KSBH.

The dynamics of a massive scalar field $\Psi$ is governed by the Klein-Gordon equation,
   \begin{equation}
		\frac{1}{\sqrt{-g}}\p_{\mu}(\sqrt{-g}g^{\mu\nu}\p_{\nu}-\mu^2)\Psi=0,
		\label{KGeq}
	\end{equation}
where $\mu$ is the field mass.
To separate the equation above, we assume the ansatz
    \begin{equation}
        \Psi=e^{i(m\varphi-\omega t)}S(\theta)R(r),
    \end{equation}
which exploits the stationarity and axisymmetry of the
background, decomposing the field into modes of frequency $\omega$ and azimuthal number $m \in \mathbb{Z}$.
The angular function $S=S_{\omega \ell m}(\theta)$ is an oblate spheroidal harmonic satisfying
  \begin{align}
		&\left(\frac{d^{2}}{d\theta^{2}}+\cot\theta\frac{d}{d\theta}\right)S \nonumber\\
		&+\left(\lambda+2ma\omega-a^{2}\omega^{2}\sin^{2}\theta-\frac{m^{2}}{\sin^{2}\theta}\right)S =0,\label{angeq}
	\end{align}
where $\lambda$ is the separation constant, determined by requiring regularity of the angular function at $\theta=0$ and $\theta=\pi$. The indices $\ell$ and $m$ label the angular modes, with $\ell$ a nonnegative integer satisfying $\ell \geq |m|$~\cite{Goldberg:1966uu,PhysRevD.73.024013}. The radial function $R=R_{\omega \ell m}(r)$ satisfies
\begin{equation}
\frac{d}{dr}\left(\Delta\frac{dR}{dr}\right)
+\left(\frac{K^2}{\Delta}-\lambda-\mu^2r(r+2d)\right)R=0,
\label{Eqrad}
\end{equation}
where
\begin{equation}
K=\omega(r^2+2 d \, r+a^2)- a m.
\label{K}
\end{equation}

\section{Quasinormal modes and degeneracies}\label{sec:3}

The QNM problem is defined by imposing purely ingoing waves at the event horizon $r_+$ and purely outgoing waves at spatial infinity~\cite{Nollert:1999ji,Konoplya:2011qq,Berti:2009kk,Bolokhov:2025uxz}. Accordingly, the radial function $R(r)$ must satisfy 
    \begin{align}
        R(r) \rightarrow 
\begin{cases} 
e^{-i(\omega - m\Omega_+)r_\star}, & r_* \rightarrow -\infty, \\
 e^{+i\kappa r_\star}, & r_* \rightarrow \infty,\label{boundC}
\end{cases}
    \end{align}
where 
\begin{equation}
        \kappa \equiv \sqrt{\omega^2-\mu^2}
        \label{kappa}
    \end{equation}
and $r_\star$ is the tortoise coordinate defined by
    \begin{equation}
        r_{\star} = \int dr\left(\frac{r(r+2d)+a^{2}}{\Delta}\right).
        \label{tortoisecoord}
    \end{equation}
For fixed parameters $(M, a, Q, \mu, \ell, m )$, the QNM boundary conditions yield a countably infinite set of frequencies $\omega_n$, labeled by the overtone number $n=0,1,2,\dots$ and typically ordered by increasing $|\operatorname{Im}(\omega_n)|$.

The QNM spectrum of the KSBH can be computed using several methods, including the continued fraction method (CFM)~\cite{Leaver:1985ax}, the Wentzel–Kramers–Brillouin (WKB) method~\cite{Konoplya2003,Konoplya2004}, and the isomonodromic method~\cite{daCunha:2021jkm}. These methods offer complementary advantages: the WKB approximation provides rapid semianalytical estimates, especially for low overtones $n$ and large $\ell$, whereas the CFM and the isomonodromic method provide high numerical accuracy. In this work, we primarily employ the CFM and crosscheck our results with the WKB method and the isomonodromic method. Details of the CFM are given below, while the WKB and isomonodromic methods are discussed in the Appendices.

\subsection{Continued fraction method}
The CFM introduced by Leaver~\cite{Leaver:1985ax} relies on a power series solution of the radial differential equation~\eqref{Eqrad}. Imposing the QNM boundary conditions leads to a recurrence relation for the series coefficients. The QNM frequencies are then determined by requiring the series to be convergent, leading to a continued fraction equation involving the recurrence coefficients~\cite{Leaver:1985ax}. 

For a massive scalar field in the KSBH spacetime, we consider the following {\it ansatz},
\begin{equation}
     \! R(r) = e^{-i \kappa r} (r-r_{-})^{\sigma_-} (r-r_+)^{\sigma_+}  \sum_{n=0}^\infty \! a_n \!\left(\frac{r-r_+}{r-r_{-}}\right)^n \! \! ,\label{Ransatz}
\end{equation}
where the exponents $\sigma_-$ and $\sigma_+$ are given by
\begin{align}
    \sigma_-&=-1+2iM\frac{(\omega+\kappa)^2}{2\kappa}+i\frac{2M\omega r_- -am}{r_+-r_-},\label{sigmap} \\
    \sigma_+ &= -i\frac{ 2M\omega r_+-a m}{r_+-r_-}.\label{sigmam}
\end{align}

 Substituting the {\it ansatz} above into Eq.~\eqref{Eqrad} yields a three-term recurrence relation for the expansion coefficients $a_n$~\cite{Kokkotas:2015uma},
    \begin{align}
    &\alpha_0^{\text{r}}a_1+\beta_0^{\text{r}}a_0=0,\label{cfeq0}\\
        &\alpha_n^{ \text{r}}a_{n+1}+\beta_n^{ \text{r}}a_n+\gamma_n^{ \text{r}}a_{n-1}=0, \qquad n\ge 1. \label{cfeq1}
    \end{align}
The recurrence coefficients $\alpha_n^{ \text{r}}$, $\beta_n^{ \text{r}}$, and $\gamma_n^{\text{r}}$  are given in Appendix \ref{appA}. Convergence of the series requires these coefficients to satisfy the continued fraction equation
    \begin{equation}
        0=\beta_0^{ \text{r}}-\frac{\alpha_0^{ \text{r}}\gamma_1^{ \text{r}}}{\beta_1^{\text{r}}-}\frac{\alpha_1^{ \text{r}}\gamma_2^{ \text{r}}}{\beta_2^{ \text{r}}-}\frac{\alpha_2^{ \text{r}}\gamma_3^{ \text{r}}}{\beta_3^{ \text{r}}-}\cdots
        \label{cfcondition}
    \end{equation}
whose roots determine the QNM frequencies $\omega$. We compute the QNM frequencies numerically by truncating the continued fraction at a sufficiently large order $N$ and solving the resulting equation using a root-finding algorithm. We validate our numerical implementation against previously published results~\cite{Konoplya:2016pmh}. The application of the CFM to the KNBH follows an analogous procedure~\cite{Konoplya:2013rxa}.

\subsection{Degeneracies in the QNM spectrum}

The QNM spectrum of BH  perturbations exhibits a rich variety of spectral phenomena, including avoided crossings and eigenvalue repulsion, in which distinct QNM branches approach one another and subsequently separate as the system parameters vary~\cite{Yang:2025dbn,PanossoMacedo:2025xnf,Rossi:2026als}. In some cases, two QNM frequencies can become exactly degenerate. An EP is a particularly important type of degeneracy at which both the QNM frequencies and their corresponding eigenfunctions coalesce~\cite{Bergholtz:2019deh,Ding:2022juv,Heiss_2012,Ashida:2020dkc}.

In the frequency domain, QNMs typically appear as simple poles of the Green's function~\cite{Leaver:1986gd}. At an EP, the coalescence of modes gives rise to a higher-order pole in the Green's function, with a two-mode degeneracy corresponding to a second-order pole~\cite{PanossoMacedo:2025xnf,Morikawa:2026fzn}. This modification of the pole structure alters the late-time behavior of perturbations and the stability properties of the spectrum under small variations of the BH parameters~\cite{Yang:2025dbn, PanossoMacedo:2025xnf, Imafuku:2026rpn, Takahashi:2026daf}. Consequently, degeneracies in the QNM spectrum leave imprints on physically observable quantities, such as the ringdown waveform and the excitation factors of the modes~\cite{Motohashi:2024fwt, Imafuku:2026rpn, Takahashi:2026daf, Zhang:2026uoo}.

The Wronskian of solutions to the radial equation~\eqref{Eqrad}, subject to the QNM boundary conditions in Eq.~\eqref{boundC}, can be used to identify degeneracies in the QNM spectrum. In particular, a degeneracy between two QNM frequencies requires the following conditions~\cite{OuldElHadj:2026vym},
\begin{align}
&W(\omega,\bm{p})|_{\rm EP}=0,\label{W}\\
&\frac{\partial W}{\partial\omega}\bigg|_{\rm EP}=0,\label{dW}
\end{align}
where the vector $\bm{p}$ represents the set of parameters characterizing the system. The first condition, Eq.~\eqref{W}, determines the ordinary QNM frequencies $\omega$ at which the Wronskian vanishes. The second condition, Eq.~\eqref{dW}, ensures that the zero is at least double, identifying a degeneracy of QNM frequencies. More generally, the coalescence of $n$ QNM frequencies requires the Wronskian and its first $n-1$ derivatives with respect to $\omega$ to vanish, yielding a zero of multiplicity at least $n$~\cite{OuldElHadj:2026vym}. The Wronskian criterion thus provides a clean and practical way to identify and classify degeneracies in the QNM spectrum of BH perturbations.

The same characterization applies to any other suitable representation of the spectral condition. Consider, for instance, a generic radial perturbation problem, such as the scalar field radial equation~\eqref{Eqrad}. The QNM boundary conditions select a discrete set of frequencies $\omega_n$ for which a nontrivial solution $R(r)$ exists. This requirement can be expressed as a secular equation,
\begin{equation}
\mathcal{F}(\omega,\boldsymbol{p}_{i})=0,
\label{eq:spectral}
\end{equation}
whose roots coincide with the QNM frequencies of the system and, equivalently, with the poles of the Green's function~\cite{Leaver:1985ax,Nollert:1992,Berti:2009}.

Different approaches may provide different representations of the secular equation \eqref{eq:spectral}, such as the continued fraction method \cite{Leaver:1985ax} or the isomonodromic method~\cite{Cavalcante:2024kmy,Cavalcante:2024swt,Cavalcante:2026vgr}. Each of these approaches yields its own form for $\mathcal{F}(\omega,\boldsymbol{p})$, and in principle any of them can be used to locate the QNM frequencies and to study their degeneracies. In the CFM, for example, one can use Eq.~\eqref{cfcondition} at $N$-th order to define $\mathcal{F}$ as
\begin{equation}
\mathcal{F}(\omega,{\bm p}) = \beta_0^{ \text{r}}-\frac{\alpha_0^{ \text{r}}\gamma_1^{ \text{r}}}{\beta_1^{ \text{r}}-}\frac{\alpha_1^{ \text{r}}\gamma_2^{ \text{r}}}{\beta_2^{ \text{r}}-}\cdots \frac{\alpha_{N-1}^{ \text{r}} \gamma_N^{ \text{r}}}{\beta_N^{ \text{r}}},
 \label{Fcf}
 \end{equation}
with the coefficients $\alpha_n^{ \text{r}}$, $\beta_n^{ \text{r}}$, and $\gamma_n^{ \text{r}}$ dependent on $(\omega, \boldsymbol{p})$. Using this representation, the conditions in Eqs.~\eqref{W} and \eqref{dW} translate into
\begin{align}
\mathcal{F}(\omega,{\bm p})|_{\rm EP}&=0,
\label{Fcfm}\\
\frac{\partial \mathcal{F}}{\partial\omega}\bigg|_{\rm EP}&=0,
\label{dFcfm}
\end{align}
and, hence, a second-order EP corresponds to a double root of the function in \eqref{Fcf}.

Equations~\eqref{Fcfm} and~\eqref{dFcfm} constitute a system of equations for the unknowns $(\omega,\bm{p})$. Solving them simultaneously yields the frequency and parameter values at which two QNMs become degenerate. Alternatively, one could compute the QNM spectrum over the entire parameter space and identify degeneracies by monitoring the frequency differences between pairs of modes. However, this approach requires a sufficiently fine grid to resolve the modes near degeneracy, and its computational cost increases with the number of parameters. Solving the conditions~\eqref{Fcfm} and~\eqref{dFcfm} directly is therefore a more efficient way to locate EPs, as it avoids the need to scan the parameter space.

The number of independent parameters in $\bm{p}$ determines the dimension of the resulting exceptional structure. For two independent parameters, Eqs.~\eqref{Fcfm} and~\eqref{dFcfm} typically yield isolated EPs. When additional parameters are introduced, the solutions can instead form exceptional lines or higher-dimensional structures. More precisely, the two complex conditions~\eqref{Fcfm} and~\eqref{dFcfm} impose four real constraints on the complex QNM frequency $\omega$ and the $L$ independent real parameters in $\bm{p}$. The solution set therefore generically has dimension $L-2$. In particular, for $L=3$ the solutions form one-dimensional curves, corresponding to exceptional lines~\cite{Cavalcante:2026vgr,Nakamoto:2026lyo,Cao:2025afs}. These exceptional lines can be traced from a known EP by varying the system parameters while maintaining the degeneracy conditions~\eqref{Fcfm} and~\eqref{dFcfm}.

A concrete example is provided in Ref.~\cite{Cavalcante:2026vgr}, where EPs in the QNM spectrum of a massive scalar field in Kerr BHs serve as initial seeds to trace exceptional lines in the KNBH parameter space. By varying the charge $Q/M$ and adjusting the remaining parameters to maintain the degeneracy conditions, the EPs are continued into the three-dimensional parameter space $\bm{p}=(Q/M,\,a/M,\,M\mu)$. The resulting curves define exceptional lines that connect the Kerr limit to configurations with nonzero electric charge.

\section{Numerical results}\label{sec:4}

\begin{table*}
\centering
\begin{tabular}{c@{\hspace{0.1cm}} c@{\hspace{0.1cm}} c @{\hspace{0.2cm}} c @{\hspace{0.2cm}} c @{\hspace{0.2cm}} c @{\hspace{0.2cm}} c}
\hhline{=======}
$Q/M$ & Method & $\ell=0$& $\ell=1$ & $\ell=2$ & $\ell=3$ & $\ell=4$\\
\hline
$0.2$ & WKB & $0.11125-0.10102i$ & $0.29491-0.09796i$ & $0.48692-0.09698i$ & $0.67993-0.09671i$ & $0.87327-0.09661i$\\
& CFM  &  $0.11124-0.10506i$ & $0.29494-0.09786i$ & $0.48692-0.09697i$ & $0.67993-0.09671i$ & $0.87327-0.09660i$\\

$0.5$ & WKB & $0.11563-0.10214i$  & $0.30621-0.09906i$ & $0.50542-0.09813i$ & $0.70571-0.09788i$ & $0.90635-0.09778i$\\
& CFM & $0.11563-0.10599i$ & $0.30622-0.09899i$ & $0.50542-0.09813i$ & $0.70571-0.09788i$ & $0.90635-0.09778i$\\

$0.7$ & WKB & $0.12136-0.10339i$ & $0.32106-0.10041i$ & $0.52976-0.09956i$ & $0.73964-0.09932i$ & $0.94990-0.09922i$ \\
& CFM & $0.12136-0.10711i$ & $0.32106-0.10037i$ & $0.52976-0.09955i$ & $0.73964-0.09932i$ & $0.94990-0.09922i$\\

$1.0$ & WKB & $0.13796-0.10442i$ & $0.36308-0.10356i$ & $0.59878-0.10291i$ & $0.83590-0.10272i$ & $1.07345-0.10264i$\\
& CFM & $0.13737-0.10962i$ & $0.36304-0.10361i$ & $0.59878-0.10291i$ & $0.83590-0.10272i$ & $1.07346-0.10264i$\\

\hhline{=======}
\end{tabular}
\caption{Comparison of the fundamental ($n=0$) QNM frequencies obtained by the sixth-order WKB method and the CFM for different values of $Q/M$ and $\ell$, when $a/M=M\mu=0$.}
\label{tableWKBCFM}
\end{table*}

In this section, we examine the massive and massless scalar QNM spectra of near-extremal KSBHs, focusing on the degeneracies that emerge as the system parameters are varied. We also compare the KSBH and KNBH spectra, computing the QNM frequencies with the CFM and cross-checking them against the isomonodromic method. We consider the $(\ell,m)=(1,1)$ and $(2,2)$ sectors and discuss the implications of these degeneracies, particularly the bifurcation of QNM families and the occurrence of hysteresis.

As a consistency check of our CFM implementation, we validate our results against those obtained using the sixth-order WKB approximation for KSBHs in the non-rotating limit. Specifically, we consider massless scalar perturbations with $a/M = M\mu = 0$, for which the effective potential has a single maximum outside the event horizon, allowing the application of the standard WKB approximation (see Appendix~\ref{appwkb}). Table~\ref{tableWKBCFM} compares the fundamental mode frequencies obtained with the two methods for $\ell = 0,\ldots,4$ and $Q/M = 0.2, 0.5, 0.7, 1.0$. For $\ell \geq 2$, the results agree to within $10^{-5}$, while larger discrepancies are observed for $\ell = 0$ and $\ell = 1$, mainly in the imaginary part. This is consistent with the known limitations of the WKB approximation, which is most accurate for $\ell \gg 1$ and becomes less reliable at low multipoles~\cite{Konoplya:2011qq}.

\begin{figure*}[t]
    \centering
    \includegraphics[width=0.48\linewidth]{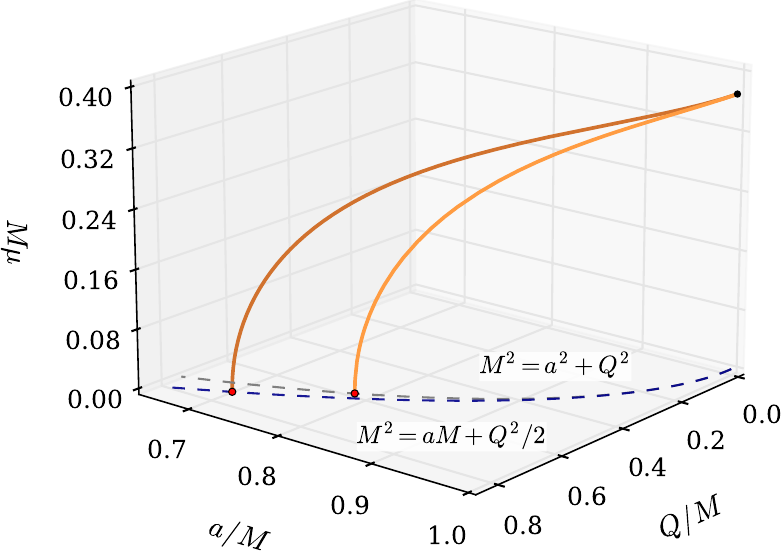}\hfill
    \includegraphics[width=0.48\linewidth]{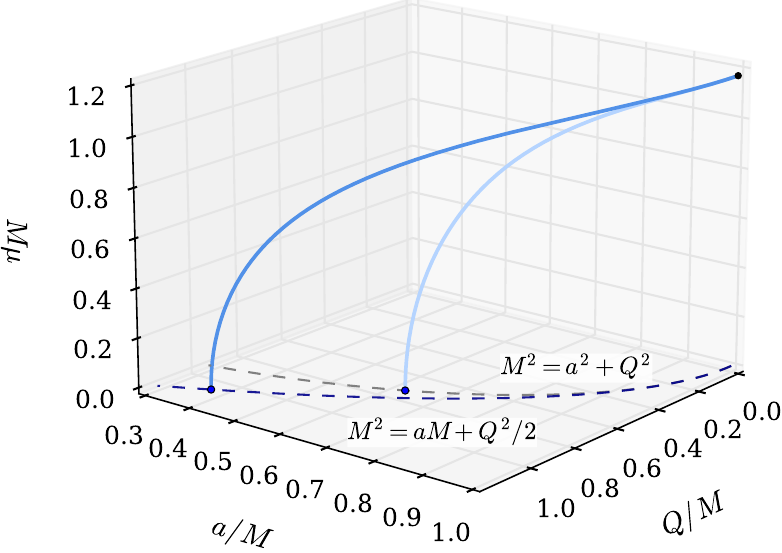}
    \caption{Exceptional lines for the $(\ell,m)=(1,1)$ and $(2,2)$ sectors of massive scalar perturbations of KNBHs and KSBHs. The light-colored (KNBHs) and dark-colored (KSBHs) curves correspond to degeneracies between the $n=0$ and $n=1$ QNM frequencies. The lines originate at the Kerr EPs in the $a/M\times M\mu$ plane (black dots) and terminate at the corresponding KN and KS EPs in the $a/M\times Q/M$ plane (red dots). The exceptional lines lie close to, but do not intersect, the extremal surfaces (cylinders) defined by $a^2+Q^2=M^2$ for KNBHs and $M^2=aM+Q^2/2$ for KSBHs. The dashed curves indicate the intersections of these surfaces with the $a/M\times Q/M$ plane, delimiting the orange and beige regions shown in Fig.~\ref{ChargeKNvsKS}.
    }
    \label{exceptionalline}
\end{figure*}

\subsection{Lines of degeneracies}
\label{subsLines}

The isolated EPs previously identified in the near-extremal Kerr QNM spectrum~\cite{Cavalcante:2024swt,Cavalcante:2024kmy,Cavalcante:2025abr} provide a starting point for investigating their extension into the parameter space of KSBHs. We trace the resulting exceptional lines as the electric charge is varied and compare our findings with those for KNBHs~\cite{Cavalcante:2026vgr}. To quantify the distance from extremality, we introduce the parameter $\delta_i$,
\begin{equation}
   \delta_i = \frac{r^{i}_+ - r^{i}_-}{2M},
   \label{extremality}
\end{equation}
where $i = \mathrm{KN}$ or $\mathrm{KS}$, with $r^{i}_-$ and $r^{i}_+$ defined in Eqs.~\eqref{rh_ks} and \eqref{rh_kn}. This parameter measures the relative separation between the outer and inner horizons and therefore provides a natural measure of proximity to extremality. The extremal limit corresponds to $\delta_i = 0$, i.e., $r^{i}_+ = r^{i}_-$, for which $M^2=a^2+Q^2$ in the KNBH case and $M^2=aM+Q^2/2$
in the KSBH case. At the other end of the range, $\delta_i=1$ corresponds to the non-rotating and uncharged configuration, $Q/M=a/M=0$. Thus, $\delta_i$ interpolates between extremal BHs and the Schwarzschild BH.

As discussed in Ref.~\cite{Cavalcante:2026vgr}, KNBHs can support exceptional lines in the massive scalar case. The same behavior emerges in the spectrum of the stringy BH, as illustrated by the dark colored curves in Fig.~\ref{exceptionalline} for the $(\ell,m)=(1,1)$ and $(2,2)$ sectors. These exceptional lines correspond to degeneracies between the $n=0$ and $n=1$ QNMs, and both originate from an EP in the Kerr limit ($Q/M=0$). We compute the exceptional lines by numerically solving Eqs.~\eqref{Fcfm} and \eqref{dFcfm} for $(a/M, M\mu, M\omega)$ at fixed values of $Q/M$, thus determining the EP coordinates in the three dimensional parameter space $(a/M, Q/M, M\mu)$. Starting from the known EPs in the Kerr limit ($Q/M=0$, black dots in Fig.~\ref{exceptionalline}), we progressively increase $Q/M$, using the EP found at each step as the initial guess for the next. As the charge increases, the scalar field mass $M\mu$ decreases, and we continue this procedure until $M\mu=0$, which determines the critical charge $(Q/M)_c$ at the endpoint of each exceptional line. The final EPs for $(\ell,m)=(1,1)$ and $(2,2)$ occur at $((a/M)_c,(Q/M)_c)\approx(0.709376,0.761782)$ and $(0.386963,1.107237)$, respectively. 

The initial and final EP coordinates along the exceptional lines shown in Fig.~\ref{exceptionalline} are listed in Table~\ref{EPs_0} for both KSBHs and KNBHs. The EPs remain close to extremality along the exceptional lines for both BH families. This is illustrated more clearly in Fig.~\ref{extremality}, which shows $\delta$ as a function of $Q/M$ along the exceptional lines for both $(\ell,m)=(1,1)$ and $(2,2)$. In particular, $\delta_{\rm KN}$ and $\delta_{\rm KS}$ remain nonzero as $M\mu\to 0$, showing that the EPs do not reach extremality in the massless limit, a feature not readily apparent in Fig.~\ref{exceptionalline}. The values of $\delta_i$ in the last column of Table~\ref{EPs_0} further show that the $(\ell,m)=(2,2)$ EPs in the massless limit (labeled Final EPs) lie closer to extremality than those for $(\ell,m)=(1,1)$. The same behavior is observed in the Kerr limit, as indicated by the critical spin values $(a/M)_c$ in the second column (Initial EPs).

\begin{table}[h]
\centering
\begin{tabular}{c  @{\hspace{0.25cm}}c  @{\hspace{0.22cm}} c  @{\hspace{0.22cm}} c  @{\hspace{0.22cm}} c  @{\hspace{0.22cm}} c @{\hspace{0.22cm}} c}
\hline \hline
\multicolumn{6}{c}{\qquad  \qquad Initial EPs {\hspace{1.0cm}} Final EPs} \\
\hline
& $(\ell,m)$  & $(M\mu)_c$ & $(a/M)_c$ & $(Q/M)_c$ & $(a/M)_c$ & $\delta_i$\\
\hline
KS & (1,1) & 0.370498 & 0.999465 & 0.761782 & 0.709376 & $0.025775$ \\
KN & (1,1) & 0.370498 & 0.999465 & 0.610458 & 0.791531 & $\ 0.028631$ \vspace{0.2cm} \\
KS & (2,2) & 1.171547 & 0.999907 & 1.107237 & 0.386963& $0.006228$ \\
KN & (2,2) & 1.171547 & 0.999907 & 0.764619 & 0.644418 & $0.009123$ \\
\hline \hline
\end{tabular}
\caption{Initial and final EPs along the exceptional lines for KSBHs and KNBHs shown in Fig.~\ref{exceptionalline}. The results correspond to the coalescence of the fundamental mode ($n=0$) and the first overtone ($n=1$) in the $(\ell,m)=(1,1)$ and $(2,2)$ sectors. The last column gives the parameter $\delta_i$, defined in Eq.~\eqref{extremality}, which measures the deviation from extremality at the final EPs ($M\mu=0$).}
\label{EPs_0}
\end{table}

On the $M\mu=0$ plane, two additional EPs were identified in the $(\ell,m)=(1,1)$ sector of KSBHs, beyond the fundamental EP discussed above. The first corresponds to the coalescence of the $(n,n+1)=(1,2)$ overtone pair and is located at $(Q/M)_1^c\approx 0.657643$ and $(a/M)_1^c\approx 0.783625$. The second, associated with the $(n,n+1)=(2,3)$ pair, occurs at $(Q/M)_2^c\approx 0.599047$ and $(a/M)_2^c\approx 0.820535$. These EPs follow a pattern similar to that reported in previous work~\cite{Cavalcante:2026vgr}: as $n$ increases, the critical charge decreases while the critical spin increases, bringing the EPs progressively closer to extremality. Although we have identified only three EPs, this behavior is consistent with a sequence accumulating towards extremality in the massless limit. Each of these EPs also marks the endpoint of an exceptional line on the $M\mu=0$ plane, analogous to those found for KNBHs~\cite{Cavalcante:2026vgr}.

\begin{figure}[h]
    \centering
    \includegraphics[width=1\linewidth]{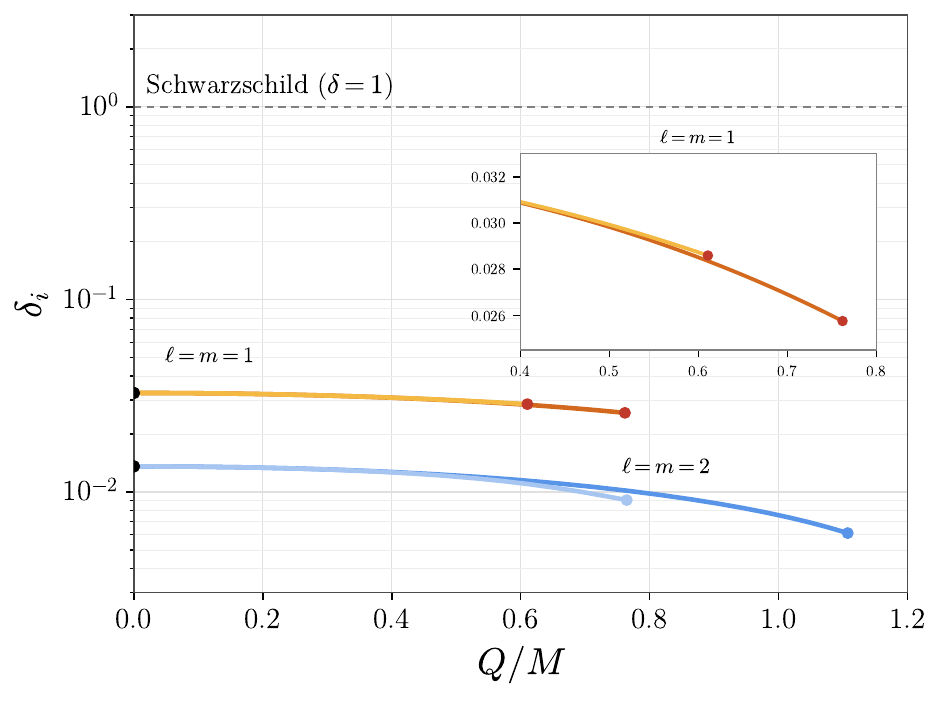}
        \caption{Near-extremal parameter $\delta_i$, with $i=\mathrm{KS},\mathrm{KN}$, as a function of $Q/M$ along the exceptional lines in the $(\ell,m)=(1,1)$ and $(2,2)$ sectors shown in Fig.~\ref{exceptionalline}. The dashed gray line indicates $\delta_i=1$, corresponding to the Schwarzschild limit ($Q/M=a/M=0$), and is included for reference. In both sectors, $\delta_i$ remains several orders of magnitude below unity, indicating that the exceptional lines lie close to extremality throughout. The inset provides a closer view of the $(\ell,m)=(1,1)$ sector, showing the small but distinct differences between the KS and KN curves.}
    \label{extremality}
\end{figure}

\subsection{Spectrum bifurcation in KSBHs}

As a BH approaches extremality, its QNM spectrum can split into two families of modes~\cite{Yang:2013uba,Kokkotas:2015uma}: ZDMs, whose damping rates vanish in the extremal limit, and DMs, whose damping rates remain finite. The EPs identified in Kerr and KN spacetimes~\cite{Cavalcante:2024kmy,Cavalcante:2024swt,Cavalcante:2026vgr} are closely related to the interplay between these two families. At these EPs, two QNM branches become degenerate, connecting modes with different near-extremal behavior. Consequently, encircling an EP can take a mode from the ZDM branch to the DM branch, or vice versa.

\begin{figure}[h]
    \centering
     \includegraphics[width=0.99
    \linewidth]{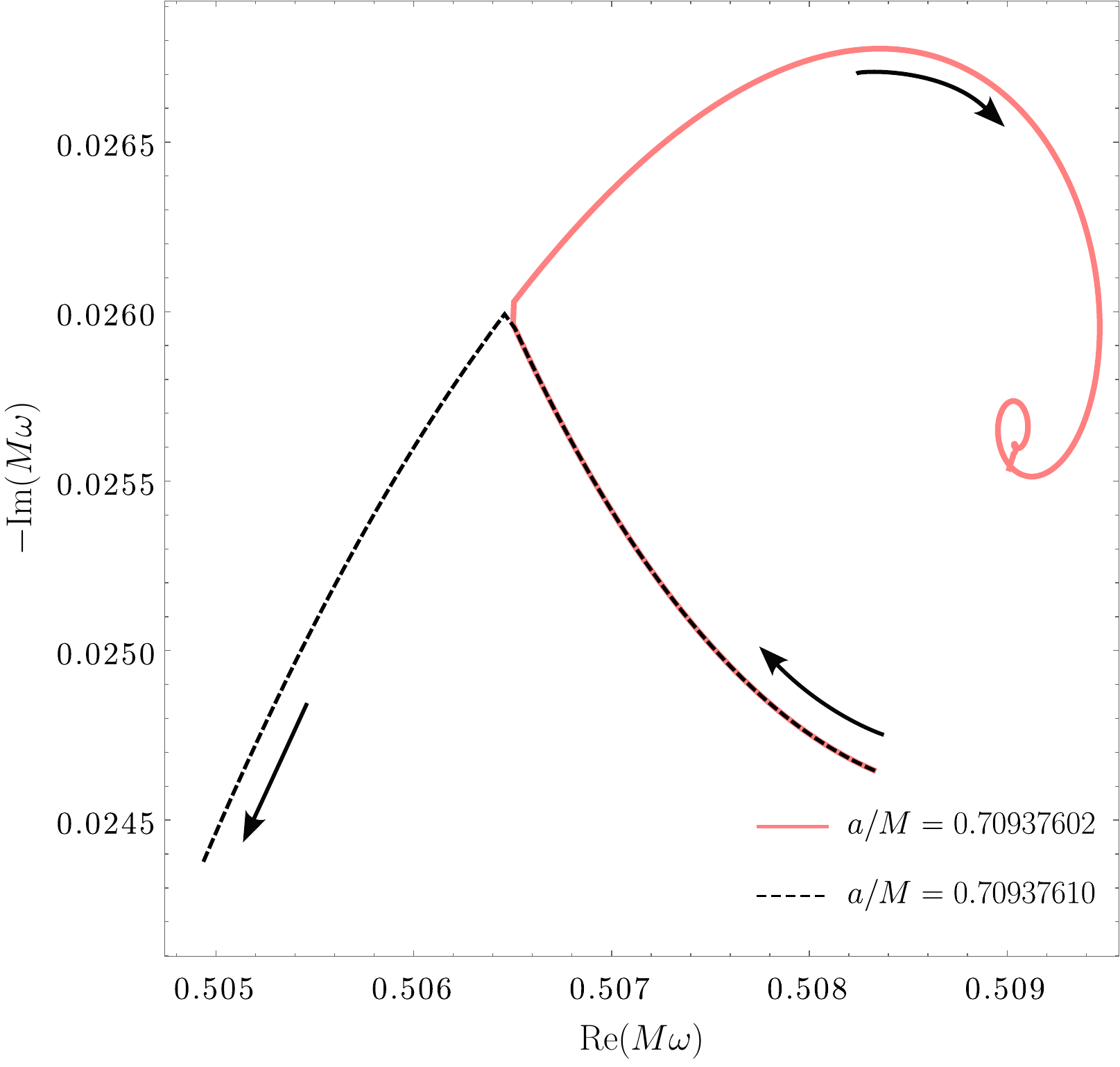}
    \caption{Fundamental QNM frequency in the $(\ell,m)=(1,1)$ sector as the electric charge $Q/M$ increases (black arrow), with $M\mu=0$. The solid red curve corresponds to a spin slightly below the critical value, $a/M\lesssim(a/M)_c$, while the dashed black curve corresponds to a spin slightly above it, $a/M\gtrsim(a/M)_c$. The two curves illustrate the change in the behavior of the fundamental mode across the EP, located at $(a/M)_c\approx0.709376$ and $(Q/M)_c\approx0.761782$.}
    \label{BandAEP}
\end{figure}

The bifurcation into DM and ZDM families was previously observed for KSBHs in Ref.~\cite{Kokkotas:2015uma}. To investigate its connection with the EPs identified in this work, we plot in Fig.~\ref{BandAEP} the fundamental QNM frequency as a function of the BH charge. The mode exhibits distinct behaviour on either side of the critical spin, $(a/M)_c\simeq 0.709376$. For $a/M \gtrsim (a/M)_c$, it belongs to the ZDM family, with $\mathrm{Im}(M\omega) \rightarrow 0$ as $Q/M$ approaches extremality. Conversely, for $a/M \lesssim (a/M)_c$, it belongs to the DM family, with $\mathrm{Im}(M\omega)$ remaining finite in the extremal limit. The branching is associated with the EP at $(a/M)_c \approx 0.709376$ and $(Q/M)_c\approx 0.761782$, where the fundamental mode coalesces with the first overtone. In fact, for $Q/M \leq (Q/M)_c$, the modes corresponding to $a/M \lesssim (a/M)_c$ and $a/M \gtrsim (a/M)_c$ are visually indistinguishable in Fig.~\ref{BandAEP}. At the critical charge, the difference between their DM and ZDM character begins to emerge. For $Q/M > (Q/M)_c$, the two branches gradually separate, making the distinction between the DM and ZDM behaviours more apparent.

\begin{figure}
    \centering
    \includegraphics[width=1.0\linewidth]{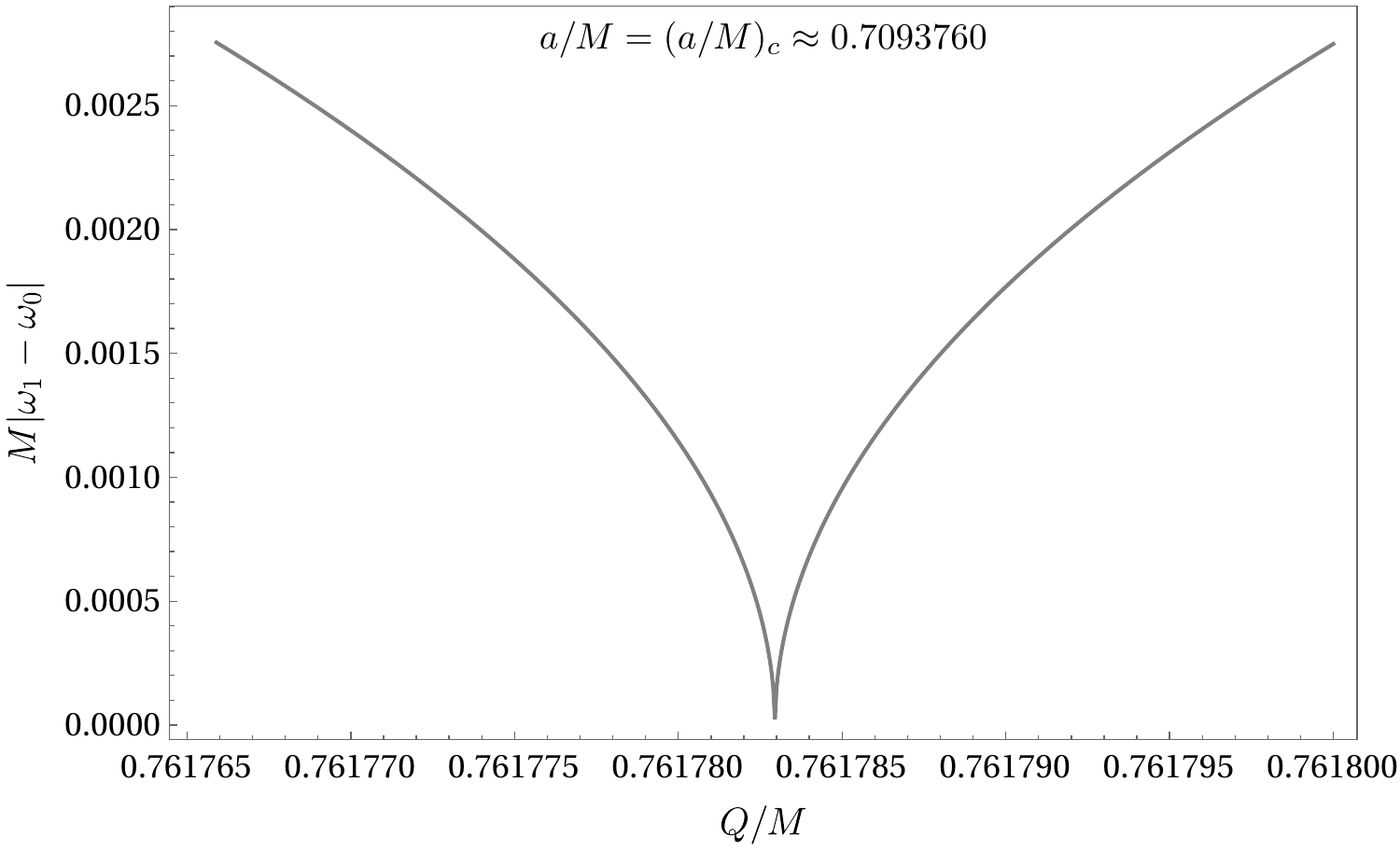}
   \caption{Absolute difference between the frequencies of the fundamental QNM ($n=0$) and the first overtone ($n=1$) as a function of the electric charge $Q/M$, at the critical spin $(a/M)_c\approx0.709376$. Along the entire curve, the condition in Eq.~\eqref{Fcfm} is satisfied. The sharp minimum occurs at the critical charge $(Q/M)_c\approx 0.761783$, where the double-root condition in Eq.~\eqref{dFcfm} is also satisfied and $M|\omega_0-\omega_1|$ vanishes, indicating the coalescence of the two QNMs.}
   \label{Absdiff}
\end{figure}

The degeneracy between the fundamental mode and the first overtone can also be illustrated by examining the absolute difference between their frequencies. Figure~\ref{Absdiff} shows $|\omega_0-\omega_1|$ as a function of $Q/M$ at the critical spin $(a/M)_c\approx0.709376$. A sharp minimum occurs at $(Q/M)_c\simeq 0.761783$, where $|\omega_0-\omega_1|$ vanishes, identifying the EP and the associated bifurcation point of the spectrum.

To complement the numerical results presented above, we derive an asymptotic expression for the frequencies of ZDMs, such as those represented by the dashed line in Fig.~\ref{BandAEP}. Following the procedure developed for KNBHs in Ref.~\cite{Cavalcante:2026vgr} and adapted here to KSBHs (see Appendix~\ref{appB}), we find that the frequency $M\omega$ admits the following expansion in the near-extremal regime:
\begin{equation}
\begin{aligned}
M\omega_k &\simeq \frac{m}{2} - \frac{i}{4}\big(1 + 2k+\sigma_0\big)\epsilon + \mathcal{O}(\epsilon^2),
\end{aligned}
\label{eq:omegEq}
\end{equation}
where $k \in \mathbb N$ indexes the ZDMs, the expansion parameter $\epsilon$ is defined by 
\begin{equation}
\cos (\epsilon) = a/(M - d), \label{epsilon}
\end{equation}
with $\sigma_0$ and $\lambda_0$ given, respectively, by Eqs.~\eqref{eq:alphaexpansion5b} and \eqref{eq:lambda0}. 
Note that $\epsilon=0$ corresponds to $a = M-d$, which is the extremal limit for KSBHs, see Eq.~\eqref{rh_ks}.

For $Q/M=0$, the expression above agrees with previous results for massive~\cite{Cavalcante:2024kmy} and massless~\cite{daCunha:2021jkm,Hod:2008zz,Casals:2019vdb} scalar fields in near extremal BH spacetimes. For $Q/M\neq0$, an analogous expression for a charged, massless scalar field was derived in Ref.~\cite{Kokkotas:2015uma}. We emphasize that the non-negative integer $k$ does not necessarily coincide with the overtone index $n$ of a given mode. This distinction arises because the spectrum near extremality contains both ZDMs and DMs, while the index $k$ labels only the ZDMs. 

\subsection{Hysteresis }

As the parameters $(a/M, Q/M, M\mu)$ are varied along a closed loop encircling at least one EP, the QNM frequencies can exchange places in the spectrum~\cite{Cavalcante:2026vgr,Cavalcante:2024swt}. For instance, the longest-lived mode may become an overtone, while an overtone becomes the fundamental mode. Consequently, after completing the loop, a given mode may not return to its original position in the spectrum, a phenomenon referred to as hysteresis in Refs.~\cite{Cavalcante:2024swt,Cavalcante:2025abr,Cavalcante:2024kmy}.

\begin{figure}
    \centering
    \includegraphics[width=0.95\linewidth]{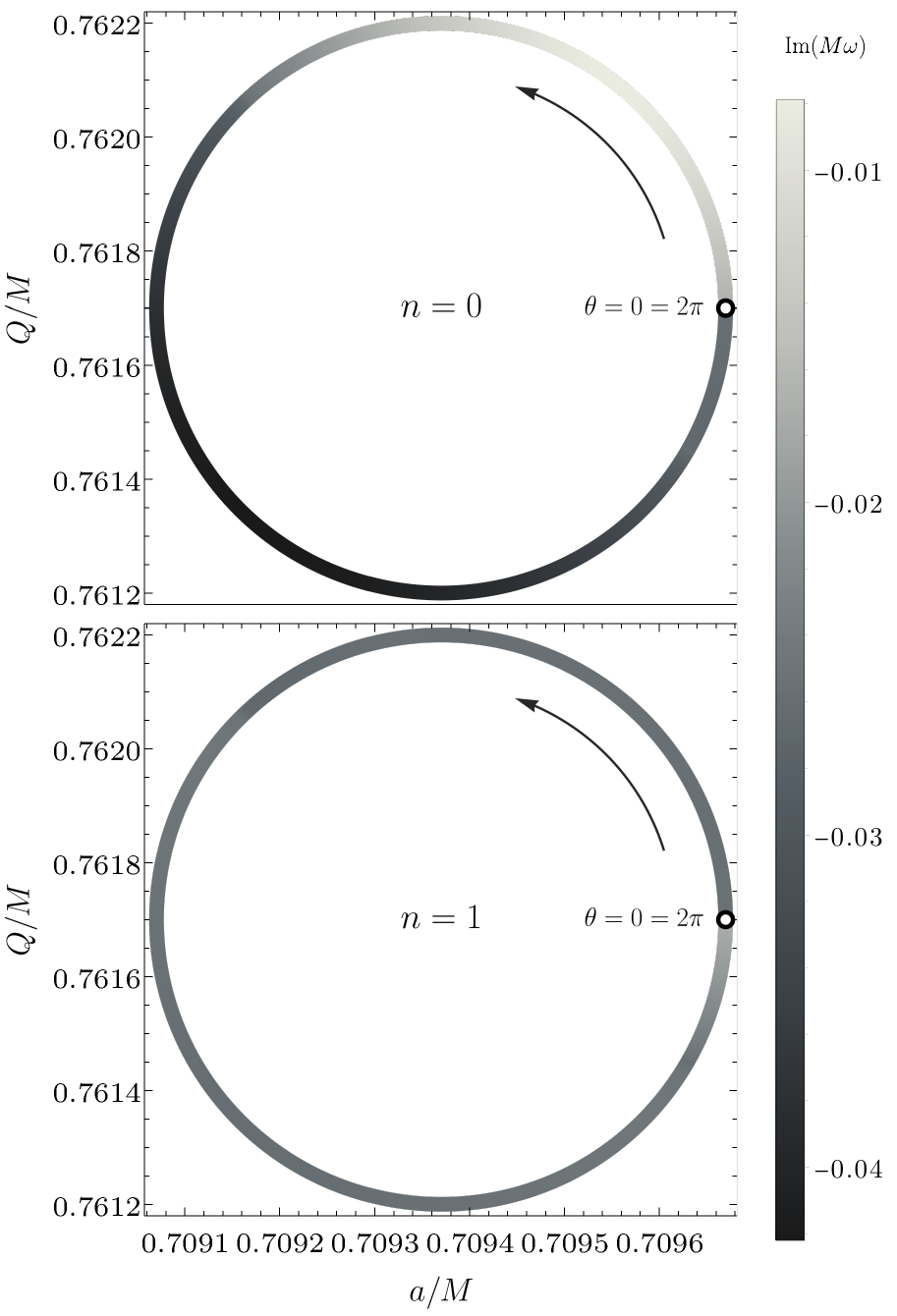}
\caption{Hysteresis around the EP in the $(\ell,m)=(1,1)$ sector for a massless scalar field. Both panels show the ellipse defined by Eqs.~\eqref{eq:contour_a} and \eqref{eq:contour_q} in the $(a/M,Q/M)$ plane, with the color indicating $\mathrm{Im}(M\omega)$ as $\theta$ increases from $0$ to $2\pi$ in the counterclockwise direction. The upper and lower panels correspond to the modes initially labeled $n=0$ and $n=1$, respectively, at $\theta=0$ (rightmost point of the ellipse). After one complete circuit, the two modes are exchanged, resulting in a discontinuity in color at the starting point.}
    \label{evolutionnn0vsn1}
\end{figure}

To illustrate this phenomenon, we consider the EP in the $(\ell,m)=(1,1)$ sector for a massless scalar field, located at $(a/M,Q/M)_c\approx(0.709376,\,0.761782)$. We define an elliptical contour  which encloses the EP in the $M\mu=0$ plane,
\begin{align}
a/M(\theta) &= 0.70937+0.0003\cos\theta, \label{eq:contour_a}\\
Q/M(\theta) &= 0.76170+0.0005\sin\theta, \label{eq:contour_q}
\end{align}
with $\theta \in [0,2\pi]$. We then follow the $n=0$ and $n=1$ QNMs adiabatically along this contour. As shown in Fig.~\ref{evolutionnn0vsn1}, the two modes exchange places after one complete revolution. The fundamental mode, initially at $M\omega_0=0.50178-0.01701i$, evolves into the first overtone at $M\omega_1=0.50935-0.02563i$, while the first overtone evolves into the fundamental mode. Thus, the modes do not return to their original positions in the spectrum after completing the loop. The exchange of the two modes after a single revolution also implies that two complete revolutions are needed to recover the original mode assignment~\cite{Berti:2025hly}. This behavior reflects the two-sheeted structure of the QNM spectrum near a second-order EP~\cite{Heiss_2012,Yang:2025dbn}.

\section{Final remarks}\label{sec:remarks}

In this work, we have investigated the presence of EPs in the QNM spectrum of a stringy BH, namely the KSBH, under massive scalar perturbations. We identified lines of degeneracy in the three-dimensional parameter space $(a/M,Q/M,M\mu)$, along which two QNMs coalesce. We also compared the exceptional structure of the KSBH spectrum with that of the KNBH, finding qualitatively similar behavior. In particular, as the electric charge increases, the EPs remain close to the extremal regime. Our results also suggest that EPs occur increasingly close to extremality for higher values of $(\ell,m)$, although a more systematic analysis of higher multipoles would be needed to establish this behavior.

To efficiently locate EPs in the QNM spectrum, we developed an approach based on the double-root nature of these degeneracies in a secular equation, represented by Eq.~\eqref{Fcf} in our case. Rather than scanning the parameter space for pairs of QNMs that approach one another, we solve the double-root conditions directly. Starting from a known EP, such as those previously identified in Kerr BHs for massive scalar perturbations, this method allows us to follow the corresponding exceptional lines as the parameters of the BH--scalar field system are varied.  The main advantage is that the exceptional lines can be traced without explicitly computing the two QNM branches that coalesce at each point of the exceptional line.

The exceptional lines identified for massive scalar perturbations extend to the $M\mu=0$ plane, where they terminate at EPs of the massless scalar spectrum. In this massless limit and for the $(\ell,m)=(1,1)$ sector, we found EPs for the overtone pairs $(n,n+1)$ with $n=0,1,2$. We also demonstrated the associated hysteresis by following the QNM frequencies along a closed path encircling an EP. We further showed that these EPs give rise to hysteresis in the QNM spectrum: after one revolution along a closed path encircling an EP, the QNM frequencies do not return to their original values, but are exchanged. This provides a direct manifestation of the nontrivial structure associated with the degeneracies in the spectrum.

Our results provide further evidence that EPs are a common feature of BH QNM spectra when sufficiently many parameters are available. A natural extension of this work is to investigate whether similar EPs and exceptional lines arise in stringy BHs for higher-spin perturbations, particularly electromagnetic and gravitational fields. Unlike test scalar perturbations, which depend only on the background geometry, these perturbations are also sensitive to the field equations of the underlying theory. In the KS spacetime, they involve couplings to the dilaton and axion fields. Comparing KS and KN BHs would therefore help determine how these additional couplings affect the exceptional structure, beyond the influence of the background geometry alone. 
More generally, studying gravitational perturbations in different BH spacetimes could help clarify the role of EPs in gravitational QNM spectra and their possible implications for BH ringdown.

\section{Data Availability}

Notebooks and supporting data for the findings presented in this work are available in Ref. \cite{xavier2026degeneracies}.
\begin{acknowledgments}
    This study was financed in part by the Coordenação de Aperfeiçoamento de Pessoal de Nível Superior - Brasil (CAPES) - Finance Code 001.
	S.~V.~M.~C.~B.~X. acknowledges the financial support from the São Paulo Research Foundation (FAPESP, Brazil), Process Number 2024/22876-0. J.~P.~C. acknowledges the financial support from the São Paulo Research Foundation (FAPESP, Brazil), Process Number 2024/15921-9.  M.~R. acknowledges partial support from the Conselho Nacional de Desenvolvimento Científico e Tecnológico (CNPq, Brazil), Grant 315991/2023-2, and from the São Paulo Research Foundation (FAPESP, Brazil), Grant 2024/00923-6.

\end{acknowledgments}

\appendix

\onecolumngrid
\section{Recurrence coefficients of the continued fraction method}\label{appA}

We provide below the coefficients of the three-term recurrence relation in Eqs.~\eqref{cfeq0} and~\eqref{cfeq1}. They are obtained by substituting the ansatz~\eqref{Ransatz}, with the exponents~\eqref{sigmap} and~\eqref{sigmam}, into the radial equation~\eqref{Eqrad}. Collecting terms with equal powers of $(r-r_+)/(r-r_-)$, we obtain
\begin{align}
\alpha_n^{\text{r}} &= (n+1)\Big(2i\big(\omega r_+(r_+ + r_- + 2d) - am\big) - (n+1)(r_+ - r_-)\Big), \label{alphan}\\
\beta_n^{ \text{r}} &= -2r_+\Omega(r_+ + r_- + 2d)\big[\omega(r_+ + r_- + 2d) + \kappa(r_+ - r_-)\big] + 2am\big[\Omega(r_+ + r_- + 2d) + (\omega+\kappa)(r_+ - r_-)\big] \nonumber\\
&\quad - i(1+2n)\Big[(r_+ + r_- + 2d)\big(\Omega(r_+ - r_-) + \omega(r_+ + r_-)\big) + \kappa(r_+ - r_-)^2 - 2am\Big] \nonumber\\
&\quad + \big(2n^2 + 2n + 1 + \lambda - \mu^2 r_+ r_-\big)(r_+ - r_-), \label{betan}\\
\gamma_n^{ \text{r}} &= -\Big(n - i\Omega(r_+ + r_- + 2d)\Big)\Big(n(r_+ - r_-) - i\big[\big(\Omega(r_+ - r_-) + 2\omega r_-\big)(r_+ + r_- + 2d) - 2am\big]\Big), \label{gamman}
\end{align}
where
\begin{equation}
 \Omega \equiv \frac{(\omega + \kappa)^2}{2\kappa},
\end{equation}
$\kappa$ is defined in Eq.~\eqref{kappa}, and $r_\pm$ are the horizons of the KSBH given in Eq.~\eqref{rh_ks}. For $\mu = 0$, one has $\kappa = \omega$ and $\Omega = 2\omega$, and Eqs.~\eqref{alphan}--\eqref{gamman} reduce to those of Ref.~\cite{Kokkotas:2015uma} for a neutral scalar field, with their parameter $b$ identified with $d$. The angular eigenvalue $\lambda$, obtained from the spheroidal equation~\eqref{angeq}, depends on the frequency and must be computed together with $\omega$ at each step of the root search. In our numerical computations, we truncate the continued fraction~\eqref{Fcf} at a depth $N$ and verify that the QNM frequencies are stable under variations in $N$ to the precision reported in the main text. Overtones with $n\geq 1$ are computed using the $n$-th inversion of the continued fraction equation~\cite{Leaver:1985ax}.

\section{WKB method}\label{appwkb}

We use the WKB method to validate our implementation of the CFM for KSBHs. The WKB method applies to wave equations of the Schr\"odinger form, with a suitable effective potential. Using the tortoise coordinate defined in Eq.~\eqref{tortoisecoord}, the radial equation~\eqref{Eqrad} can be written as
\begin{equation}
\left[\frac{d^{2}}{dr_{\star}^{2}}+(\omega^2 - V_{\omega\ell m})\right]R(r_{\star})=0,
\label{RWeq}
\end{equation}
where $V_{\omega\ell m}$ is the effective potential.
In general, the effective potential $V_{\omega\ell m}$ is not well suited to the standard WKB approximation. For $a/M \neq 0$, it depends on the frequency through $K$ and $\lambda$ (see Eq.~\eqref{K}), while for $M\mu \neq 0$, the asymptotic behavior changes and a potential well may form.

\begin{figure}
    \centering
    \includegraphics[width=0.7\linewidth]{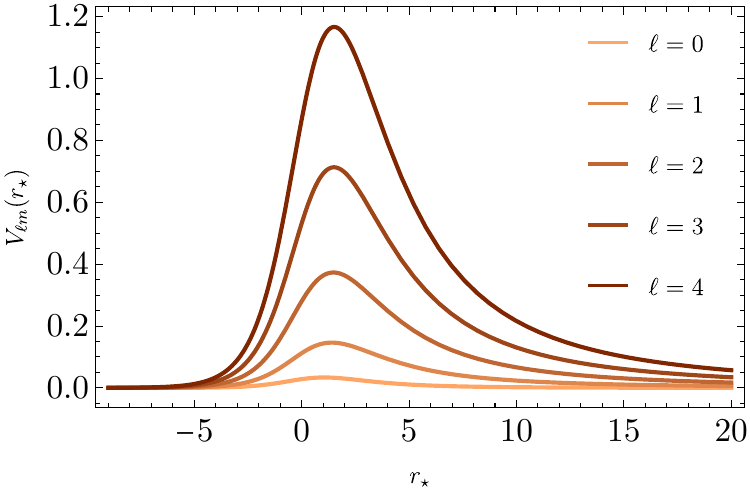}
    \caption{Effective potential $V_{\ell m}$ of the KSBH, as a function of the tortoise coordinate $r_\star$, for $Q/M = 1$ and $a/M = M\mu = 0$, for different values of $\ell$.}
    \label{wplot}
\end{figure}

We therefore restrict the comparison to $a/M = M\mu = 0$, which is the Gibbons–Maeda–Garfinkle–Horowitz–Strominger BH~\cite{Gibbons:1987ps,Garfinkle:1990qj}. In this limit, the only departure from the Schwarzschild potential comes from the charge, through $d$, allowing us to test the KS specific part of our implementation. The angular eigenvalue reduces to $\lambda = \ell(\ell+1)$, and the explicit expression for the effective potential can be obtained from Eq.~(23) of Ref.~\cite{Xavier:2023ljy} by setting $a/M = 0$, yielding
\begin{equation}
\begin{split}
     V_{\ell m} &= \frac{\left[ Q^2+M(r-2M) \right]}{4r^2(Q^2+Mr)^4} \Biggl\{ -Q^6 - Q^4 \Bigl[ 2M^2  - Mr + 4Mr\ell(\ell+1) \Bigr] + Q^2 \Bigl[ 4M^3r + 8M^2r^2\ell(\ell+1) \Bigr] + \\
    & + 8M^4r^2 + 4M^3r^3\ell(\ell+1) \Biggr\}
\end{split}
\label{VwlmWKB}
\end{equation}
Note that this effective potential is independent of the frequency. As shown in Fig.~\ref{wplot}, it vanishes at the horizon and at spatial infinity, with a single maximum at $r_\star=r_\star^0$. This is the barrier structure required for the standard WKB treatment.

The WKB method determines the QNM frequencies by matching approximate solutions of Eq.~\eqref{RWeq} in three regions separated by the two turning points, where $\omega ^ 2 - V_{\ell m} = 0$~\cite{SchutzWill1985,IyerWill1987}. Region I extends from the outer turning point to $r_\star \to +\infty$, while region III extends from the inner turning point to $r_\star \to -\infty$. In these regions, we use the WKB ansatz
\begin{equation}
    R \sim \exp\left( \sum_{j=0}^\infty \frac{S_j\, h^j}{h}\right),
    \label{wkbansatz}
\end{equation}
where $h$ is a formal parameter used to keep track of the order of the approximation. The functions $S_j$ are determined by substituting this expansion into Eq.~\eqref{RWeq} and collecting terms of the same order in $h$. In each region, the expansion yields two independent solutions, $R_+$ and $R_-$, propagating toward increasing and decreasing $r_\star$, respectively. The general solutions can then be written as
\begin{align}
    R &\sim Z_{\rm in}^{\rm I}\, R_-^{\rm I} + Z_{\rm out}^{\rm I}\, R_+^{\rm I}, && \text{region I}, \label{solutionI}\\
    R &\sim Z_{\rm in}^{\rm III}\, R_+^{\rm III} + Z_{\rm out}^{\rm III}\, R_-^{\rm III}, && \text{region III}, \label{solutionIII}
\end{align}
where the labels ``in'' and ``out'' refer to waves propagating toward and away from the potential barrier, respectively.

The solutions in regions I and III are connected through region II, located between the turning points. When $\omega^2$ is close to the maximum of the potential barrier, the turning points lie close to each other, and $V_{\ell m}$ can be expanded around $r_\star^0$. Keeping terms up to second order in $r_\star-r_\star^0$, Eq.~\eqref{RWeq} reduces to the parabolic cylinder equation, whose general solution is
\begin{equation}
    R = A\, D_{\nu}(t) + B\, D_{-\nu-1}(it),
    \label{solutionII}
\end{equation}
where $D_\nu$ denotes the parabolic cylinder function, $t$ is a rescaled coordinate proportional to $r_\star-r_\star^0$, and
\begin{equation}
    \nu + \frac{1}{2} = \left. \frac{i\left( \omega^2- \,V_{\ell m}\right)}{\sqrt{-2V_{\ell m}''}} \right|_{r_\star^0},
    \label{wkbeq}
\end{equation}
with primes denoting derivatives with respect to $r_\star$. Matching the solution in region II to those in regions I and III relates the coefficients in Eqs.~\eqref{solutionI} and \eqref{solutionIII}. The QNM boundary conditions of Sec.~\ref{sec:3} require that no waves be incident on the barrier from either side, i.e., $Z_{\rm in}^{\rm I} = Z_{\rm in}^{\rm III} = 0$. This gives $\nu = n$, where $n$ is a non-negative integer identified with the overtone number, recovering the first-order WKB formula of Ref.~\cite{SchutzWill1985}.

Including higher-order corrections in the WKB matching procedure leads to the generalized condition~\cite{IyerWill1987,Konoplya2003,Konoplya2004}
\begin{equation}
     n + \frac{1}{2}=\left. \frac{i\left( \omega^2-\,V_{\ell m}\right)}{\sqrt{-2V_{\ell m}''}} \right|_{r_\star^0} - \sum_{k=2}^{6}\Lambda_k,
    \label{wkbeq6}
\end{equation}
where the corrections $\Lambda_k$ depend on $n$ and on derivatives of $V_{\ell m}$ at $r_0$ up to order $2k$. The terms $\Lambda_2$ and $\Lambda_3$ were derived in Ref.~\cite{IyerWill1987}, extending the approximation to third order, while $\Lambda_4$ to $\Lambda_6$ were obtained in Refs.~\cite{Konoplya2003,Konoplya2004}, extending it to sixth order. The WKB approximation was subsequently extended to 13th order in Ref.~\cite{MatyjasekOpala}. We use the sixth-order expression to compute the WKB frequencies presented in Table~\ref{tableWKBCFM}. We remark that the WKB expansion is asymptotic, so increasing the order does not necessarily improve the accuracy, although it often does in practice~\cite{Konoplya:2011qq,KonoplyaZhidenkoZinhailo}. The approximation is most reliable for $\ell \gg n$, explaining the larger discrepancies observed for $\ell = 0$ and $\ell = 1$ in Table~\ref{tableWKBCFM}.

\section{Isomonodromic method and the analytic expression for ZDMs of Near-Extremal KSBHs}\label{appB}

In this appendix, we derive an approximate expression for the ZDM frequencies of near-extremal KSBHs, considering the general case $\ell=m\geq 1$. We follow the isomonodromic approach of Ref.~\cite{daCunha:2021jkm}, which relates the QNM problem to the monodromy properties of the radial differential equation. These properties describe how the solutions transform under analytic continuation around the singular points of the equation. The local monodromy data are characterized by the parameters $\theta_0$, $\theta_t$, and $\theta_\star$, associated with the inner horizon, outer horizon, and spatial infinity, respectively. For the KSBH, one has
\begin{equation}
    \theta_0 =\frac{2 i(a^2 \omega -a m + r_- (2 d + r_-) \omega)}{r_+ - r_-}, \qquad \theta_{t}= 2\sigma_+, \qquad \theta_{\star}= \frac{2M\mu^2 - 2 (2 d + 2M)\omega^2}{\kappa},
    \label{eq:param}
\end{equation}
with $r_\pm$ the horizons of the KSBH given in Eq.~\eqref{rh_ks}, $\kappa$ defined in Eq.~\eqref{kappa}, and $\sigma_+$ defined in Eq.~\eqref{sigmam}. In addition to these local monodromy data, the method involves two further quantities: the composite monodromy parameter $\sigma$, which characterizes the monodromy associated with both horizons, and the parameter $\eta$, which encodes the connection data between the corresponding local solutions.
The isomonodromic method relates the coefficients of the radial equation to the monodromy data through the Painlev\'e-V tau function, denoted by $\tau_V$, see Refs.~\cite{daCunha:2021jkm,Cavalcante:2023rdy} for further details. This function depends on the monodromy parameters and on the deformation parameter $t$, which is proportional to the separation between the horizons,
\begin{equation}
t=2i\kappa(r_+ - r_-).
\end{equation}
The QNM boundary conditions impose a relation among the monodromy data, while the expansion of $\tau_V$ relates these data to the coefficients of the radial equation.

To obtain an asymptotic expression for the ZDM frequencies, we expand $M\omega$, the angular eigenvalue $\lambda$, and the  monodromy parameter $\sigma$ in powers of the small parameter $\epsilon$ defined in Eq.~\eqref{epsilon}. We write
\begin{equation}\label{eq:exppar3}
M\omega = \beta_0 + \beta_1\epsilon + \mathcal{O}(\epsilon^2), \qquad
\lambda = \lambda_0 + \lambda_1 \epsilon + \mathcal{O}(\epsilon^2), \qquad
\sigma = 1 + \sigma_0 + \sigma_1 \epsilon + \mathcal{O}(\epsilon^2),
\end{equation}
where $\beta_0=m/2$. To determine $\beta_1$, we follow the procedure of Ref.~\cite{daCunha:2021jkm}, starting from the small-$t$ expansion of $\tau_V$. Inverting this expansion leads to the relation
\begin{equation} \label{piV}
\Pi_V\, e^{i\pi\eta}\, t^{\sigma-1} = \chi_V(t),
\end{equation}
where $0<\operatorname{Re}(\sigma)<1$, $\Pi_V$ is a function of the monodromy parameters and $\chi_V(t)$ admits the perturbative expansion
\begin{equation}
    \chi_V(t) = \sum_{i=0} c_i\, t^i.
\end{equation}
The leading terms of the asymptotic expansion for $\chi_V(t)$ are found to be
\begin{equation}
\chi_V(t)=1+
\left[
\frac{\theta_\star(\theta_0^2-\theta_t^2)}{4\sigma^2}
-\frac{\theta_\star(\theta_0^2-\theta_t^2)}{4(\sigma-2)^2}
\right]t+\mathcal{O}(t^2).
\label{eq}
\end{equation}
The functions $e^{i\pi\eta}$ and $\Pi_V$, on the other hand, can be expressed in terms of sine and gamma functions as
\begin{equation}
  e^{i\pi\eta}
  = e^{-i\pi\sigma}\frac{\sin\tfrac{\pi}{2}(\theta_\star+\sigma)}{
    \sin\tfrac{\pi}{2}(\theta_\star-\sigma)}\prod_{\zeta=\pm1}
  \frac{\sin\tfrac{\pi}{2}(f_{+}(\zeta))
    }{
    \sin\tfrac{\pi}{2}(f_{-}(\zeta))
    }, \quad \qquad  \Pi_V = \frac{\Gamma^2(2-\sigma)}{\Gamma^2(\sigma)}
\frac{\Gamma(\tfrac{1}{2}(\theta_\star+\sigma))}{
	\Gamma(1+\tfrac{1}{2}(\theta_\star-\sigma))} \prod_{\zeta=\pm 1}
  \frac{\Gamma(\frac{1}{2}f_{+}(\zeta))}{
    \Gamma(1+\frac{1}{2}f_{-}(\zeta))},
  \label{eq:quantizationV}
\end{equation}
where $f_{\pm}(\zeta)=\theta_{t}+\zeta\theta_0\pm\sigma$.
In the near-extremal limit, $r_+ - r_- \to 0$, and therefore $t\to 0$, justifying the use of the small-$t$ expansion.

Substituting Eqs.~\eqref{eq:quantizationV} into Eq.~\eqref{piV}, using expansions of \eqref{eq:param}, and including the first terms for the expansion of $\chi_V(t)$, we find that the lowest-order terms in $\epsilon$ in Eq.~\eqref{piV} yield
\begin{equation}
\begin{aligned}
      e^{-\pi i\sigma_0}\frac{\Gamma(1-\sigma_0)^2}{
    \Gamma(1+\sigma_0)^2}
  \frac{\Gamma(\tfrac{1}{2}(1+\sigma_0-4i \beta_1))}{
    \Gamma(\tfrac{1}{2}(1-\sigma_0-4i \beta_1))}
  \frac{\Gamma(\tfrac{1}{2}(1+\sigma_0-2im)}{
    \Gamma(\tfrac{1}{2}(1-\sigma_0-2im))}&
  \frac{\Gamma(\tfrac{1}{2}(1+\sigma_0-\gamma))}{
    \Gamma(\tfrac{1}{2}(1-\sigma_0-\gamma))}\times \\ &\times
    \bigg(4(1-d')\epsilon\sqrt{M^2\mu^2-\beta_0^2}\bigg)^{\sigma_0} =
    1+{\cal O}({\epsilon},\epsilon\log\epsilon)
\end{aligned}
  \label{eq:quantbeta}
\end{equation}
where
\begin{equation}\label{eq:gamma}
\gamma = \frac{2(m^2-2M^2\mu^2-2d'M^2\mu^2)}
{\sqrt{4M^2\mu^2-m^2}},
\end{equation}
with $d'=d/M = Q^2/M^2$. In the Kerr limit, corresponding to $d'=0$ (equivalently, $Q=0$), Eqs.~\eqref{eq:quantbeta} and \eqref{eq:gamma} reproduce Eqs.~(55) and (56) in Ref.~\cite{Cavalcante:2024kmy}, respectively. Note that even though the expansion of $\chi_V$ is analytic in $\epsilon$, the term ${t}^{\sigma-1}$ in Eq.~\eqref{piV} introduces non-analytic terms, such as $\epsilon\log \epsilon$. As we take the limit $\epsilon\rightarrow 0$ in Eq.~\eqref{eq:quantbeta}, the term $\epsilon^{\sigma_0}$ approaches zero if Re$(\sigma_0)>0$. In that case, the only way to satisfy Eq.~\eqref{eq:quantbeta} is if the argument of one of the gamma functions in the numerator is a non-positive integer, which we denote as $-k$, with $k \in \mathbb{Z}_+$. In fact, we can show that Eq.~\eqref{eq:quantbeta} is satisfied, to the lowest order in $\epsilon$, if
\begin{equation}
\beta_1=-\frac{i}{4}(2k+1+\sigma_0).
\end{equation}

In order to compute $\sigma_0$, we use the expansion for the accessory parameter given in Eq. (26) of Ref. \cite{daCunha:2021jkm}. Using this expansion and considering Eqs.~\eqref{eq:exppar3} and \eqref{eq:param}, we arrive at the following expression for $\sigma_0$:
\begin{equation}\label{eq:alphaexpansion5b}
\sigma_0 =
\pm \sqrt{1 + (d' (2 + d')-7) m^2 + 4\lambda_0 + 4(d'-1)^2M^2\mu^2}
\end{equation}
where $\lambda_0$ denotes the leading-order term in the expansion of the angular eigenvalue near extremality in Eq.~\eqref{eq:exppar3}. This term is given by 
\begin{equation} \label{eq:lambda0}
\lambda_0 = 
\ell(1+\ell) - \frac{1}{4}(d'-1)^{2}\left(m^{2}-4M^{2}\mu^{2}\right)
+ \frac{(d'-1)^{2}\left(\ell(1+\ell)+m^{2}-1\right)\left(m^{2}-4M^{2}\mu^{2}\right)}{8\ell(1+\ell)-6}.
\end{equation}
For $\ell=m=1$, with $d'=0$, this expression reproduces the corresponding result in Ref.~\cite{Cavalcante:2025abr}.

Finally, substituting the expressions for $\beta_1$ and $\sigma_0$ into the perturbative expression \eqref{eq:exppar3} for $M\omega$, we find that the near extremal frequencies of the ZDMs, indexed by the integer $k$, are approximated by
\begin{equation}
\begin{aligned}
  M\omega_{k} \simeq & \frac{m}{2} - i \frac{(k + 1/2 + \sigma_0)}{2} \epsilon +\mathcal{O}(\epsilon^2). \label{eq:omegaeqover}
  \end{aligned}
\end{equation}

\twocolumngrid
\bibliography{ref}
\end{document}